\documentclass[aps,letterpaper,twocolumn,pra,footinbib,floatfix,showpacs]{revtex4}
\usepackage{amsmath}
\usepackage{amsfonts}
\usepackage{amssymb}
\usepackage[scanall]{psfrag}
\usepackage{psfrag}
\usepackage{graphicx}
\usepackage{color}

\begin{document}
\newcommand{\of}[1]{\left( #1 \right)}
\newcommand{\sqof}[1]{\left[ #1 \right]}
\newcommand{\abs}[1]{\left| #1 \right|}
\newcommand{\avg}[1]{\left< #1 \right>}
\newcommand{\cuof}[1]{\left \{ #1 \right \} }
\newcommand{\pil}{\frac{\pi}{L}}
\newcommand{\bx}{\mathbf{x}}
\newcommand{\by}{\mathbf{y}}
\newcommand{\bk}{\mathbf{k}}
\newcommand{\bp}{\mathbf{p}}
\newcommand{\bl}{\mathbf{l}}
\newcommand{\bq}{\mathbf{q}}
\newcommand{\bs}{\mathbf{s}}
\newcommand{\psibar}{\overline{\psi}}
\newcommand{\svec}{\overrightarrow{\sigma}}
\newcommand{\dvec}{\overrightarrow{\partial}}
\newcommand{\bA}{\mathbf{A}}
\newcommand{\bdelta}{\mathbf{\delta}}
\newcommand{\bK}{\mathbf{K}}
\newcommand{\bQ}{\mathbf{Q}}
\newcommand{\bG}{\mathbf{G}}
\newcommand{\bw}{\mathbf{w}}
\newcommand{\up}{\uparrow}
\newcommand{\down}{\downarrow}
\author{Eliot Kapit}
\email{ek259@cornell.edu}
\author{Erich J. Mueller}
\email{em256@cornell.edu}
\affiliation{Laboratory of Atomic and Solid State Physics, Cornell University}
\title{Optical Lattice Hamiltonians for Relativistic Quantum Electrodynamics}

\begin{abstract}
We show how interpenetrating optical lattices containing Bose-Fermi mixtures can be constructed to emulate the thermodynamics of quantum electrodynamics (QED). We present models of neutral atoms on lattices in 1+1, 2+1 and 3+1 dimensions whose low energy effective action reduces to that of photons coupled to Dirac fermions of the corresponding dimensionality. We give special attention to 2+1 dimensional electrodynamics (QED3) and discuss how two of its most interesting features, chiral symmetry breaking and Chern-Simons physics, could be observed experimentally. 

\end{abstract}
\maketitle

\section{Introduction}

Planar quantum electrodynamics (QED3) has been of great theoretical interest for many years. The restriction to lower dimensionality profoundly changes the structure of the electromagnetic field and its symmetries \cite{deserreview}. In two spatial dimensions (2+1d) the electromagnetic potential between point charges that scales as $\log r$ instead of $r^{-1}$, leading one to expect charge confinement in the appropriate limit. QED3 with massive fermions is particularly rich. One of the two possible Lorentz invariant mass terms breaks parity and time reversal symmetry, leading to a topological Chern-Simons term in the photon action. This term produces a gap in the photon spectrum and makes the effective interaction between the fermions short ranged. The other Lorentz invariant fermion mass term breaks ``chiral" symmetry and is spontaneously generated at low temperatues. Both confinement and chiral symmetry breaking occur in quantum chromodynamics;  since QED3 is a much simpler theory it has been studied as a model system for these effects \cite{appelquist1,appelquist2,aitchisonqed3,fischer,dagotto,appelquist3}. Further, QED3 has been proposed as an effective theory for spin-liquids \cite{hermele1,hermele2} and the pseudogap phase of high-temperature superconductors \cite{aitchison,kim,franz,rantner}. A direct experimental realization of QED3 could therefore be relevant to many areas of both condensed matter and high energy physics.

Recent advances in cold atom physics and optical lattices \cite{bloch} have made the simulation of QED3 a real experimental possibility. Dirac fermions appear at half filling in a hexagonal optical lattice \cite{zhao,shao,becker,soltanpanahi,poletti}, mimicking the relativistic band structure of graphene \cite{neto}. They also appear in a square lattice with appropriate complex hopping matrix elements \cite{liusquare}. This has led to a number of proposals to emulate relativistic physics with cold atoms \cite{osterloh,zhu,angelakis,cirac,liusquare,yu}. Here we extend these previous proposals by showing how to emulate QED3 with cold atoms. By combining mixtures of bosons and fermions on interpenetrating optical lattices with appropriately tuned densities and interactions, both relativistic fermions and the full 3-component dynamical gauge field can be constructed, allowing QED3 to be probed directly. The linear Bogoliubov excitations of three species of bosonic atoms correspond to the three components of the vector field $A_\mu$. One of the boson species occupies the same lattice as the fermions, while the other two sit on alternating bonds between the fermion sites. These bosons modulate the fermion hopping amplitudes, leading to an interaction that at low energy has the form of a gauge coupling $e \psibar \of{ \gamma_\mu A_\mu } \psi$. Being a lattice model, higher energy terms in this theory will naturally deviate from the emergent Lorentz and gauge invariances, but at low energies, these terms are heavily suppressed, and after applying the Wilsonian prescription of discarding irrelevant operators, the resulting Hamiltonian is that of QED3.

It is important to note that in the real time formalism, the time ($A_0$) component of the gauge field is canonically quantized by operators with negative norm, with causality and Hermiticity preserved in real states by the Ward identity \cite{peskin}. This is a general feature of vector bosons, and cannot be duplicated with cold atoms. However, in the finite temperature (imaginary time) formalism, all the components of the $A$ field have positive norm. Thus, we focus on producing a quantum simulator which reproduces the thermodynamics of QED3. Many of the most interesting properties of QED3 (such as mass generation and symmetry breaking) can be probed through equilibrium thermodynamics. Further, some real-time correlations can be accessed through the fluctuation-dissipation theorem.

Our construction is very different from traditional examples of gauge fields in condensed matter systems \cite{hermele1,hermele2,aitchison,kim,franz,rantner,wenlee}. 
Typically, these gauge theories arise from using descriptions with  redundant degrees of freedom (such as slave bosons in the large $U$ Fermi Hubbard model). The complicated underlying physics of these emergent gauge theories makes it difficult to use them as simulators of gauge physics itself. The approach that we take is different; instead of formulating a redundant description of a strongly interacting system, we take a weakly interacting system with a large number of degrees of freedom and tune the couplings and particle densities so that the low energy physics is identical to thermal QED. 

To construct our theory, we necessarily choose a particular gauge. The Hamiltonian of our theory is identical to that of QED in the Feynman gauge. Our theory is therefore not a true gauge theory in the formal sense of being described by an equivalence class of distinct actions and wavefunctions; rather, it is a theory of a relativistic fermion interacting with three neutral, independent massless bosons through a coupling of the form $\psibar \gamma_\mu A_\mu \psi$. All physical observables of QED are gauge independent, and despite its fixed gauge, our simulator should correctly reproduce all such observables.

The remainder of the paper is organized in the following manner. In section \ref{sech2d}, we will describe the canonical quantization of QED3 in the Feynman gauge, taking care to denote the various terms in the Hamiltonian that must be replicated in our cold atom system. In section \ref{secsli}, we will then outline our model and show that its low-energy limit is identical to the Hamiltonian formulation of QED3. At the end of that section we will describe some of the more dramatic properties of QED3 and how they would manifest themselves in cold atom experiments. Finally, in sections~\ref{sec1d} through \ref{seccon} we will offer concluding remarks and discuss extensions of our theory to 1+1 and 3+1 dimensions.

\section{QED and Gauge Freedom In 2+1 Dimensions}\label{sech2d}

Our goal is to simulate the Euclidean action for 2+1 dimensional quantum electrodynamics,
\begin{eqnarray}\label{LE}
L_{E} =  i \psibar \gamma_{\mu} \partial_{\mu} \psi + e \psibar \of{ \gamma_\mu A_\mu } \psi - \frac{1}{4} F_{\mu \nu} F_{\mu \nu}.
\end{eqnarray}
We adopt the conventions that spacetime indicies which run from 0 to 2 are represented by Greek letters, and objects in boldface will represent the two spatial components of the Lorentz 3-vectors. Our imaginary time metric and $\gamma$ matricies are
\begin{eqnarray}\label{gammas}
g_{\mu \nu} = - \delta_{\mu \nu}, \psibar = i \psi^{\dagger} \gamma_0, \nonumber \\
\gamma_0 = i \sigma_z,  \gamma_1 = i \sigma_x,  \gamma_2 = i \sigma_y.
\end{eqnarray}
and $\psi$ is a two-component fermionic spinor. The $\sigma_{xyz}$ are Pauli matricies, and $F_{\mu \nu} = \partial_{\mu} A_{\nu} - \partial_{\nu} A_{\mu} $ is the electromagnetic field strength tensor. Summation is implied for all repeated indicies, and we have abandoned the standard convention of raising and lowering indicies to emphasize that our imaginary time metric is trivial. In section III we will generalize this action to massive fermions. We work in the imaginary time formalism \cite{freedman,dolanjackiw} to ensure that the photon propagator $D_{\mu \nu}$ is positive definite and therefore all photon states have positive norm and there are no ghostlike degrees of freedom. We note that in a non-Abelian theory, negative norm degrees of freedom will be found even in the imaginary time formalism. Thus, our approach is not appropriate for simulating QCD.

We will simulate this Hamiltonian by designing an optical lattice system that exhibits identical degrees of freedom and Feynman rules up to lowest order in powers of momenta. To arrive at a simple enough form of the action for quantum simulation, we will employ the Faddeev-Popov gauge fixing procedure \cite{peskin}. This scheme enforces the gauge constraint
\begin{eqnarray}\label{fp1}
\partial_\mu A_\mu = w \of{x}
\end{eqnarray}
and then path integrates over $w$ to remove the gauge degrees of freedom from the Lagrangian. So long as we restrict ourselves to calculating only gauge-invariant quantities, the integration over $w$ can be performed trivially. This alters the Euclidean action by adding a term of the form
\begin{eqnarray}\label{fp2}
L_{E} \rightarrow L_{E} - \frac{1}{2 \xi} \of{\partial_\mu A_\mu}^{2},
\end{eqnarray}
where $\xi$ is an arbitrary parameter. If we then choose $\xi = 1$ (the Feynman gauge), the off-diagonal terms in the photon propagator cancel out and we are left with the action
\begin{eqnarray}
L_{E} = i \psibar \gamma_{\mu} \partial_{\mu} \psi + e \psibar \of{ \gamma_\mu A_\mu } \psi + \frac{1}{2} \partial_\mu A_\nu \partial_\mu A_\nu.
\end{eqnarray}
Within this gauge, our system is described by the Hamiltonian
\begin{eqnarray}\label{H}
H &=&  v_F \psibar \of{ \svec \cdot \bk} \psi + e \psibar \of{\sigma_z A_0 + i \svec \cdot \bA} \psi \nonumber \\ & &+ \frac{1}{2} \partial_{\nu} A_{\mu} \partial_{\nu} A_{\mu},  \\ 
& & \frac{1}{2} \partial_{\nu} A_{\mu} \partial_{\nu} A_{\mu} = \sum_{r=0}^{2} \sum_{\bk}  \omega_\bk \alpha_{r \bk}^\dagger \alpha_{r \bk} \nonumber
\end{eqnarray}
The $\alpha_{r}$ operators annihilate a photon of polarization $r$, and $\omega_\bk = v_s \abs{k}$; this is simply the canonical quantization of the free photon Hamiltonian. Note that, unlike in Minkowski space, $\sqof{\alpha_{r \bk}, \alpha_{s \bp}^{\dagger} }= \delta_{rs} \delta^{2} \of{\bk-\bp}$ describes only positive norm states, since the negative time-time component of the photon propagator is made positive by the rotation to imaginary time. The gauge field $A$ is quantized in terms of these bosons by
\begin{eqnarray}\label{A}
A_\mu \of{\bx} = \sum_{r=0}^{2} \sum_{\bk} \frac{f_{r \bk \mu}}{\sqrt{2 \omega_\bk} } \sqof{ \alpha_{r \bk}  e^{i \bk \cdot \bx} + \alpha_{r \bk}^{\dagger} e^{-i \bk \cdot \bx} }.
\end{eqnarray}
Here, the $f_{r \bk \mu}$ are real, 3-element polarization vectors satisfying $f_{r \bk} \cdot f_{s \bk} = \delta_{rs}$. The Ward identity \cite{aitchisontext} implies that real photons are transverse, ie $\bk \cdot \mathbf{f}_{s \bk} = 0$ for all physical states; only transverse photons will be produced or destroyed in observable processes. The momentum-space propagators for the fermion and gauge field are
\begin{eqnarray}
D_{F} = \frac{-1}{\gamma_{\mu} k^{\mu} + m +i \epsilon}, \\ D_{ \mu \nu} = \frac{- \delta_{\mu \nu}}{k^{2} + i \epsilon}. \nonumber
\end{eqnarray}
We will now show how cold atoms can be used to directly implement the Hamiltonian (\ref{H}).

\section{2+1 Dimensional Optical Lattice QED}\label{secsli}

\begin{figure}
\psfrag{a1}{$a_{1}$}
\psfrag{a2}{$a_{2}$}
\psfrag{i1}{$i$}
\psfrag{i2}{$-i$}
\includegraphics[width=3in]{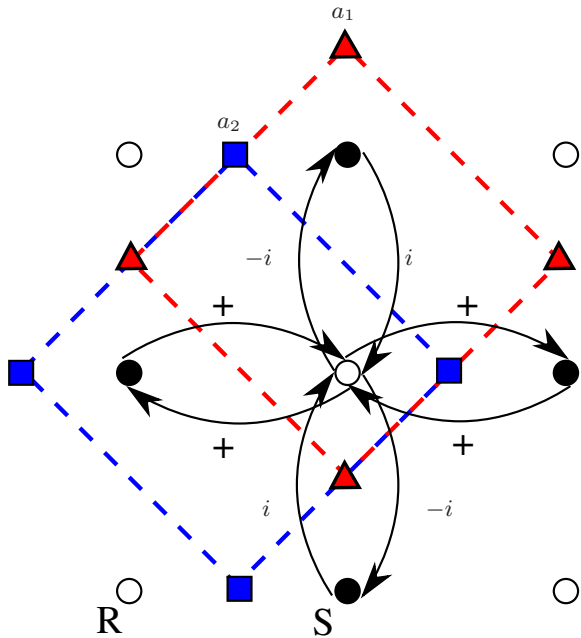}

\caption{(Color online) Square lattice for QED3. The fermions are confined to the lattice sites $R$ and $S$ denoted by white and black circles. The bosons sit on the red triangles and blue squares denoted by $\alpha_{1}$ and $\alpha_{2}$. Overall phases for fermion hopping are labeled in the figure.}\label{sqlat}
\end{figure}

We give two alternative geometries for the lattice confining our fermions. We require that the system exhibits ``Dirac points" instead of a Fermi surface; all zero- or low-energy fermionic excitations  must occur near some finite set of points $\bK$ in momentum space, and have the dispersion relation $\epsilon_\bp = \pm \sqrt{v_{F}^{2} \bp^2 + m^2 }$ for small $\bp$ measured with respect to $\bK$. The most well-known system of this type is the hexagonal lattice \cite{zhao,shao,becker,soltanpanahi,poletti}, realized in condensed matter in graphene \cite{neto} and carbon nanotubes. In cold atoms, the ability to directly manipulate the phases of lattice hopping amplitudes enables the creation of Dirac points in a square lattice \cite{liusquare,yu}, something that hasn't been realized in materials. We describe both approaches.

\subsection{Square Lattice Implementation}

Our square lattice approach is based on the geometry proposed by Liu \textit{et al} \cite{liusquare} to study the anomalous quantum Hall effect. Consider noninteracting fermions in an square optical lattice with nearest neighbor hopping $t$ and lattice spacing $l$. We add to the hopping amplitude a set of phases such that
\begin{eqnarray}\label{Hnn}
H &=& - \sum_{jk} t_{jk} \psi_{j}^{\dagger} \psi_{k}; \\ t_{jk} &=& t \; \cuof{y_{j} = y_{k}  }\mathrm{(horizontal \, hop)}, \nonumber \\ &=& i t \of{-1}^{x_{j}+y_{j} } \; \cuof{y_{j} \neq y_{k} } (\mathrm{vertical \, hop}) . \nonumber
\end{eqnarray}
The operators $\psi_{j}$ and $\psi_{j}^{\dagger}$ destroy or create a fermion at site $j$. This structure is mathematically identical to a constant magnetic field with flux density $\pi$ (or half of a fundamental magnetic flux quantum $\phi_0$) per plaquette. This effective magnetic field could be realized through a spatially dependent Raman coupling to an external laser field \cite{spielman1,spielman2,spielman3,jaksch}.

To find the band structure, we break the lattice into two sublattices $R$ and $S$, depending on whether $x_j + y_j$ is even or odd (figure \ref{sqlat}). The momentum space Hamiltonian can be written as the dot product of a pair of two-component spinors through a matrix. This matrix $M$ has no diagonal elements because nearest-neighbor hopping always transfers a fermion between sublattices,
\begin{eqnarray}
H_{D} = -2t\sum_{\bk} \cuof{\psi_{R \bk}^{\dagger},\psi_{S \bk}^{\dagger} }.M.\cuof{\psi_{R \bk},\psi_{S \bk} }, \\
M = \of{ \begin{array}{cc} 0 & \cos k_x l + i \cos k_y l \\ \cos k_x l - i \cos k_y l & 0 \end{array}}. \nonumber
\end{eqnarray}
This Hamiltonian produces a Dirac cone dispersion about the four $\bK$ points $\bK_{\pm \pm} = \cuof{\pm \pi/2l, \pm \pi/2l}$. Only two of these points ($\bK_{++}$ and $\bK_{+-}$) are distinct; the others are related to them by symmetry. We predominantly focus on $\bK_{++} = \pi/2l \cuof{1,1} = \bK$ and let $\bk = \bK + \bp$; the structure at the other $\bK$ point is the same up to the signs of the Dirac $\gamma$ matricies. With the definitions (\ref{gammas}), small $\bp$ and $\psi_\bp = \cuof{\psi_{R \bp}, \psi_{S \bp}}$, our Hamiltonian becomes
\begin{eqnarray}
H_{D} &=& 2 t l \sum_\bp \psi_{\bp}^{\dagger} \of{\begin{array}{cc} 0 & p_x + i p_y \\  p_x  - i p_y  & 0 \end{array}} \psi_\bp  \\ &=& 2 t l \sum_\bp \psibar_\bp \of{ \gamma \cdot \bp } \psi_\bp, \nonumber 
\end{eqnarray}
We now add a gauge field to the theory by introducing three species of lattice bosons. One of the three boson species ($a_0$) lives on the same lattice as the fermions and the other two ($a_1$ and $a_2$) occupy interpenetrating square lattices as shown in figure \ref{sqlat}. If the interaction between the bosons and fermions is repulsive, the presence of $a_1$ bosons between a pair of neighboring fermion lattice sites acts as an effective potential barrier that reduces the hopping amplitude between the two sites. Similarly, the presence of the fermions will modulate the boson hopping and lead to a nearest-neighbor repulsion term between bosons and fermions. The total Hamiltonian incorporating all of these effects is
\begin{widetext}
\begin{eqnarray}\label{Htotal}
& & \cuof{\bx \in L_{f}, \, \by_{b=1,2} \in L_{b=1,2}} \nonumber \\
H &=& -t \sum_{\bx,\bs} \psi_{R/S \bx}^{\dagger} \psi_{S/R \bx \pm \bs} e^{i \phi_{\bx,\bs}} + \sum_{ b; \bk} \epsilon_{b \bk} a_{b \bk}^{\dagger} a_{b \bk} + \sum_{\bx; b} \frac{V_{b}}{2} n_{b \by_{b}} \of{n_{b \by_{b}} - 1} + g_0 \sum_{\bx} \psi_{\bx}^{\dagger} \psi_{\bx} n_{0 \bx}  \\  & &+  \sum_{b=1,2; \bx, \bs} g_{b} \psi_{R/S\bx}^{\dagger} \psi_{S/R\bx \pm \bs} n_{b \by_{b} = \bx \pm \frac{\bs}{2}}
+\sum_{b;\bx} g_{b} n_{b \by_{b}} \psi_{R/S \bx}^{\dagger} \psi_{R/S \bx} + \sum_{b=1,2; \by} g_{b}' a_{b \by_{b} \pm 2 \bs}^{\dagger} a_{b \by_{b}} \psi_{R/S \bx = \by_{b} \pm \frac{\bs}{2}}^{\dagger} \psi_{R/S \bx = \by_{b} \pm \frac{\bs}{2}}. \nonumber
\end{eqnarray}
\end{widetext}
The sum over $b$ is over boson flavors (either 0,1,2 or 1,2 as specified), the vector $\bs$ joins nearest neighbors and is equal to $\of{\pm l \widehat{x}, 0}, \of{0, \pm l \widehat{y}}$, where, as previously defined, $l$ is the lattice spacing. $L_f$ is the fermion lattice, and $L_b$ is the lattice associated with species $b$. The operator $n_{b \by}$ measures the density of $a_b$ particles at $\by$. From left to right, the terms on the first line are the fermion hopping term, the boson kinetic terms $\epsilon_{b \bk}$ (which include the boson chemical potentials), the weak repulsive contact interaction $V_b$ between bosons, and the local repulsion between fermions and the $a_0$ bosons. The terms on the second line are the boson mediation of fermion hopping, nearest neighbor Bose-Fermi repulsion, and fermion mediation of boson hopping. We assume that there are no direct interactions between bosons of different species as they are spatially separated. The $g_{b}$ are the effective interactions between the bosons and fermions, which can be tuned by manipulating the optical potential depth or through Feshbach resonances and are calculated from the overlap of the neighboring Wannier functions \cite{albus}. For simplicity, we shall assume that $g_b = g_{b}'$, a situation that we expect can be reached through appropriate fine tuning. In the tight-binding limit, next nearest neighbor hopping and any additional interactions are vanishingly small. The last two terms in Eq. (\ref{Htotal}) can lead to unwanted terms in our low energy action. As calculated below, their contributions cancel each other if the $b=1,2$ bosons are condensed about $\bK = \cuof{\pi/2l,\pi/2l}$. This can be engineered, for example, by using the techniques of Ref.~\cite{spielman2} to introduce phases on the hopping matrix elements. Under these conditions,
\begin{eqnarray}
\epsilon_{b=1,2 \bk} &=& +2 t_{b=1,2} \of{ \cos \frac{\bk_{x}}{2 l} +  \cos \frac{\bk_{y}}{2 l} } - \mu_{b}, \\
\epsilon_{0 \bk} &=& -2 t_0  \of{ \cos \frac{\bk_{x}}{ l} +  \cos \frac{\bk_{y}}{ l} } - \mu_{0}, \nonumber
\end{eqnarray}
where the $t_b$ are positive numbers parametrizing the boson hopping. 
We shall now show that, with appropriate choices for the various parameters, the low energy excitations of eq. (\ref{Htotal}) are exactly those of QED3.

We first assume that the temperature is low enough that all of the boson fields have condensed; if this is the case, we can expand the Bose operators around the condensate expectation values, dropping terms that are quadratic or higher in boson fluctuations. For example, if $b=1,2$
\begin{eqnarray}\label{bogexp}
n_{b \bx} &=& n_b + \sqrt{n_b} \of{-1}^{x+y} \of{\widehat{a}_{b \bx}^{\dagger} + \widehat{a}_{b \bx}} + O \of{\frac{1}{n} } \\ &\equiv& n_b + \sqrt{n_b} \of{-1}^{x+y} G_{b \bx}, \nonumber
\end{eqnarray}
where $\widehat{a}_{b \bx} = a_{b \bx} - \sqrt{n_{b}}$. To treat the bosons' self-interactions and vacuum expectation value, we perform a Bogoliubov transformation \cite{wen}. For a given species this process gives 
\begin{eqnarray}\label{bog}
H_{b} &=& \sum_{\bk} E_{b\bk} \alpha_{b \bk}^{\dagger} \alpha_{b \bk}, \nonumber \\
E_{b\bk} &=& \sqrt{\epsilon_{b \bk} \of{\epsilon_{b\bk} + 2 n_b V_b} }, \nonumber \\
\widehat{a}_{b \bk} &=& u_{b \bk} \alpha_{b \bk} - v_{b \bk} \alpha_{b -\bk}^{\dagger}. \nonumber
\end{eqnarray} 
Here $u_{b \bk}$ and $v_{b \bk}$ are real constants satisfying $u_{\bk}^{2} - v_{\bk}^{2} =1$,
\begin{eqnarray}\label{uv}
u_{b \bk} &=& \sqrt{ \frac{\epsilon_{b\bk} + n_b V_b}{E_{b\bk}}    + \frac{1}{2} }, \\ v_{b \bk} &=& -\sqrt{ \frac{\epsilon_{b\bk} + n_b V_b}{E_{\bk_b}}    - \frac{1}{2} }. \nonumber
\end{eqnarray}
Furthermore, at low momenta these coherence factors simplify,
\begin{equation}
\lim_{\bk \to 0} u_{\bk} = - \lim_{\bk \to 0} v_{\bk} = \sqrt{n V/E_{\bk}}, 
\end{equation}
allowing us to use the low-energy approximation
\begin{equation}\label{Gk}
G_{b \bk} \approx 2  \sqrt{n_{b} \frac{V}{E_{\bk}}} \sqof{ \alpha_{b \bk} + \alpha_{b -\bk}^{\dagger}}. 
\end{equation}
The ``speed of light" for these bosons is $v_{b} = \sqrt{n_b V_b/m_b}$. We shall tune the $t_{b}$ so that the Fermi velocity $v_{F}$ matches each of the three $v_{b}$.

Now consider the first term on the second line of Eq. (\ref{Htotal}). Expanding the fields in momentum space and summing over $\bk$ and $b$ yeilds
\begin{eqnarray}\label{gxy}
& &\sum_{b=1,2; \bx}g_{b} \psi_{R/S\bx}^{\dagger} \psi_{S/R\bx \pm \bl} n_{b \bx \pm \bl_b/2} =  \\
 & &+ \sum_{b=1,2,\bk,\bp,\bq} \sqrt{n_b} g_{b} \of{G_{b \bk} \psibar_{\bp}^{\dagger} \gamma_b \psi_{\bq} } \delta_{\bk+\bp-\bq} \nonumber \\
 & &+ \sum_{b=1,2;\bk}g_b n_{b} \of{\psibar_{ \bk}^{\dagger} \gamma_b \psi_{\bk} } + O \of{\frac{1}{n},l\bk,l\bp,l\bq}. \nonumber
\end{eqnarray}
Here, the momenta in $G$ and $\psi$ are expanded about $\cuof{\pi/2l,\pi/2l}$ and $\bK$, respectively. Similarly, the local repulsion term adds two terms of the form
\begin{eqnarray}\label{glocal}
& &g_0 \sum_{\bx} \psi_{\bx}^{\dagger} \psi_{\bx} n_{0 \bx} = n_0 g_0 \sum_{\bk} \of{\psibar_{\bk} \gamma_0 \psi_\bk}  \\
& &+ g_0 \sqrt{n_0}  \sum_{\bk,\bp,\bq} \of{G_{0 \bk} \psibar_{\bp}^{\dagger} \gamma_0 \psi_{\bq} } \delta_{\bk+\bp-\bq} \nonumber \\ & &+ O \of{\frac{1}{n},l\bk,l\bp,l\bq}. \nonumber
\end{eqnarray}
We can account for the remaining two terms on the second line of (\ref{Htotal}) through a similar expansion. Assuming that $g_{b} = g_{b}'$, we find that with our choice of boson dispersion (with a minumum at $\cuof{\pi/2l,\pi/2l}$)  these two terms precisely cancel each other to lowest order in $\bk$. No such cancellation occurs if the bosons are condensed about $\bk=0$. The resulting low energy theory is
\begin{widetext}
\begin{eqnarray}\label{Hreduced}
H =  v_F \sum_\bk \psibar_\bk  \gamma \cdot \bk \psi_\bk  + \sum_{b=0}^{2} \sum_{\bk}  \omega_\bk \alpha_{b \bk}^\dagger \alpha_{b \bk} + \sum_{b=0}^{2} \sum_{\bk,\bp} g_{b}  \of{\sqrt{n_b} G_{b \bk} + n_b }\psibar_\bp \gamma_b \psi_{\bk+\bp} + O \of{\frac{1}{n},l\bk,l\bp,l\bq}.
\end{eqnarray}
\end{widetext}
Comparing $G_{b \bk}$ in (\ref{Gk}) to the canonical quantization of the gauge field $A$ in (\ref{A}) leads us to write
\begin{eqnarray}
G_{b \bx} = 2 \sqrt{2 n_b V_{b} } A_{b \bx}.
\end{eqnarray}
Provided that $v_{b} = \sqrt{n_b V_b/m_b} = v_F$ and $2 n_b \sqrt{2 V_b} g_b = e$ for all three boson species, we finally see that (\ref{Hreduced}) reduces to (\ref{H}) with the addition of a constant shift of the gauge
\begin{eqnarray}
\psibar \gamma_\mu A_\mu \psi \to \psibar \gamma_\mu \of{A_\mu + B_\mu } \psi, \quad B_\mu = \frac{1}{\sqrt{V_b}}.
\end{eqnarray}
This constant shift has no effect on the overall physics of the theory and simply shifts the location of the Dirac points in $\bk$ space. 

For $K^{40}$ fermions and $Rb^{87}$ fermions with a lattice depth of $4 E_{R}$  (where $E_R$ is the recoil energy) for each lattice, an average boson density of unity, and a Bose-Fermi scattering length of $-177 a_0$ \cite{ferlaino}, the coupling constant $e$ (in units of square root temperature) is approximately $2 \of{\mathrm{nK}}^{1/2}$. This value 
can be tuned considerably through Feshbach resonances, optical lattice tuning, and adjusting the boson filling.

\subsection{Hexagonal lattice}

\begin{figure}
\psfrag{a1}{$a_{1}$}
\psfrag{a2}{$a_{2}$}
\psfrag{b}{$(a_1,a_2)$}
\includegraphics[width=3in]{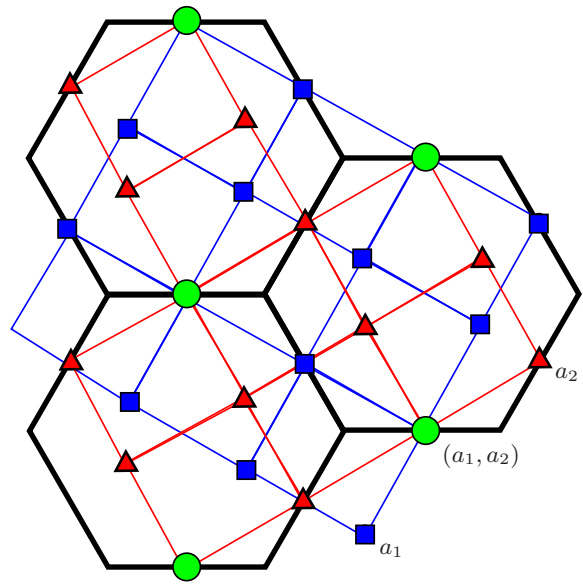}
\caption{(Color online) Hexagonal optical lattice for QED3. The fermions and $a_{0}$ bosons sit on the verticies of the hexagonal lattice (in black). The two other boson species $a_{1}$ and $a_{2}$ are on the interpenetrating orthorhombic lattices; $a_{1}$ bosons are confined to the blue squares, $a_{2}$ bosons are confined to the red triangle), and the green circles indicate sites shared by the two lattices. The hopping parameters of these lattices can be tuned so that the low-energy dispersion $\epsilon_{\bk}$ is simply $\bk^{2}/2 m^{*}$ for each of the three species.}\label{hexlat}
\end{figure}

Our theory can also be realized on a hexagonal optical lattice \cite{neto,zhao,shao,becker,soltanpanahi,poletti}. Assume that the spacing between verticies in the hexagons is $l$. The basis vectors $\delta$ connecting a pair of nearest neighbor sites are
\begin{eqnarray}
\bdelta_{1} = \of{\frac{-l}{\sqrt{3}},0}, \bdelta_{2} = \of{\frac{l}{2\sqrt{3}}, - \frac{l}{2} }, \bdelta_{3} = \of{\frac{l}{2\sqrt{3}},  \frac{l}{2} }.
\end{eqnarray}
After dividing the lattice into two triangular sublattices $R$ and $S$ (see figure \ref{hexlat}) we arrive at the fermion hopping Hamiltonian
\begin{eqnarray}
H = - t \sum_\bk \sum_{j=1}^{3} \sqof{ e^{i \bk \cdot \bdelta_j} \psi_{R \bk}^{\dagger} \psi_{S \bk} + h.c. }.
\end{eqnarray}
The coefficients $\sum_{j=1} e^{i \bk \cdot \bdelta_j}$ vanish at six points $\mathbf{K}$, of which only two are unique -- the other four are related by reciprocal lattice vectors. We shall choose $\mathbf{K}_{\pm} = \pm \frac{2 \pi}{l} \of{0,\frac{2}{3}}$. For small $\bk$ defined relative to one of these two $\mathbf{K}$ points, the Hamiltonian is the familiar Dirac cone:
\begin{eqnarray}
H_{D} = \frac{t l \sqrt{3}}{2} i \psibar \svec \cdot \bk \psi.
\end{eqnarray}
As in the previous section, $\psi = \cuof{\psi_{R},\psi_{S}}$ is a two-component spinor and $\psibar = \psi^{\dagger} \gamma_{0}$. We again construct the gauge field by adding three BEC fields, as shown in figure (\ref{hexlat}). The boson $a_{0}$ sits on the same lattice as the fermions, but $a_{1}$ and $a_{2}$ are instead on the links between the fermion lattice sites. Specifically, $a_{1}$ is chosen to occupy sites separated by $\delta_1$ and $\delta_2$ and $a_{2}$ occupies sites separated by $\delta_1$ and $\delta_3$. This structure breaks the $C_{6}$ symmetry of the lattice, but if the coefficients are chosen properly the low energy theory will be rotationally invariant. As in the square lattice case, we require that the hopping matrix elements for $b_{1}$ and $b_{2}$ to be opposite in sign to those of $b_{0}$. This will again allow the unwanted nearest neighbor repulsion and fermion mediation of boson hopping to cancel each other. Ignoring these terms, we're left with:
\begin{widetext}
\begin{eqnarray}\label{H12hex}
H_{int} = \sum_{\bx} \of{g_{\delta_{1}} \psi_{R \bx}^{\dagger} \psi_{S \bx+\delta_{1}}  \of{n_{1 \bx + \frac{\delta_{1}}{2}}+n_{2 \bx+\frac{\delta_{1}}{2}}} + g_{\delta_{2}} \psi_{R \bx}^{\dagger} \psi_{S \bx+\delta_{2} } n_{1 \bx\frac{\delta_{2}}{2}} + g_{\delta_{3}} \psi_{R \bx}^{\dagger} \psi_{S \bx+\delta_{3} } n_{2 \bx\frac{\delta_{3}}{2}} + h.c.}.
\end{eqnarray}
\end{widetext}
The couplings $g$ are determined from the fermion-boson scattering amplitudes and from the geometric position of the bosons relative to the fermion lattice sites \footnote{For example, one could reduce these amplitudes by staggering some of the boson lattice sites in the $z$ direction out of the plane.}. Repeating the steps in equations (\ref{bogexp}-\ref{Hreduced}), and setting 
\begin{eqnarray}
g_{\delta_{2}} = g_{\delta_{3}} = g, g_{\delta_{1}} = \frac{1}{2} \of{1+\sqrt{3}} g.
\end{eqnarray}
yeilds the QED Hamiltonian (\ref{H}) with an effective charge $e = 2 \sqrt{3 V} n g$ and a speed of light $v_{F} = t l \sqrt{3}/2$.

\subsection{Observable Consequences}

\subsubsection{Physical and Virtual Photons}

An important property of QED3 is the fact that there is only a single physical polarization for the photon.
This means that in any experiment neither time-like nor longitudinal photons can be created or destroyed.  Consequently, in our emulation of QED3, no perturbation of the fermions will produce these ``unphysical" photons.  This is a dramatic observable.

\subsubsection{Spontaneous Symmetry Breaking}\label{ssb}

At sufficiently low temperatures, thermal QED3 is expected to undergo a phase transition to a state with broken chiral symmetry and a finite fermion mass \cite{appelquist1,appelquist2,aitchisonqed3,fischer,dagotto,appelquist3,templeton,bashir,feng}. 
Chiral symmetry relates the two Dirac points of our lattice model, and as we will show below, chiral symmetry breaking on our square lattice corresponds to a commensurate $\cuof{\pi,\pi}$ charge density wave.

To define the Chiral symmetry, we combine the fermions at the two Dirac points into a four-component Dirac spinor.  Defining  $\bk = \bK_{\pm} + \bp$, we write 
%
\begin{eqnarray}\label{rep}
H &=& i \psibar \of{\Gamma \cdot \bp}, \psi \\
\psi_{\bp} &=& \cuof{\psi_{R \bK_{+} +\bp}, \psi_{S \bK_{+} +\bp}, \psi_{R \bK_{-} +\bp}, \psi_{S \bK_{-} +\bp} }, \nonumber \\
\Gamma_{0} &=& \of{ \begin{array}{cc}
\sigma_{z} & 0 \\
0 & \sigma_{z} \end{array}}, \Gamma_{x} = \of{ \begin{array}{cc}
\sigma_{x} & 0 \\
0 & \sigma_{x} \end{array}}, \nonumber \\ \Gamma_{y} &=& \of{ \begin{array}{cc}
\sigma_{y} & 0 \\
0 & -\sigma_{y} \end{array}}. \nonumber
\end{eqnarray}
The photon coupling becomes $e  \psibar \Gamma_\mu A_\mu \psi$, and we have simply obtained a larger representation of the Dirac algebra. This theory 
is invarient under transformations of the form
\begin{eqnarray}
\psi \rightarrow \exp \of{i \epsilon_{1} C_{1} } \exp \of{ i \epsilon_{2} C_{2} } \psi, 
\end{eqnarray}
for any real constants $\epsilon_{1}$ and $\epsilon_{2}$ and $4 \times 4$ chiral matricies $C_{1}$ and $C_{2}$ defined by
\begin{eqnarray}
C_{1} = \of{ \begin{array}{cc}
0 & \sigma_{y} \\ \sigma_{y} & 0 \end{array}}, C_{2} = i \of{ \begin{array}{cc}
0 & -\sigma_{y} \\ \sigma_{y} & 0 \end{array}}. 
\end{eqnarray}
These matricies anticommute with each other and with all three $\Gamma$ matricies. They are the generators of a symmetry transformation that simultaneously flips between the two sublattices and the two $\bK_{\pm}$ points.

The low temperature phase of QED3 breaks this symmetry, with $\langle  \sum_k \bar \psi_k\psi_k\rangle\neq 0$.  In our real space lattice this order parameter is a charge density  wave of the form
\begin{equation}
\langle  \sum_k \bar \psi_k\psi_k\rangle=\sum_{\bx \in L_{S}} \psi_{S \bx}^{\dagger} \psi_{S \bx} - \sum_{\bx \in L_{R}} \psi_{R \bx}^{\dagger} \psi_{R \bx}.
\end{equation}
This order can be observed using light scattering \cite{bragg} or through single-site imaging of the fermions.  Unfortunately the transition temperature for this spontaneous symmetry breaking is $T_c\sim 10^{-2} e^{2}$ \cite{dagotto,he}.  Using our previous estimate $e \sim 3 (\mbox{nK})^{1/2}$, we find that $T_{c} \sim 100 \mbox{pK}$ is beyond the reach of current experiments.

\begin{figure}
\includegraphics[width=3in]{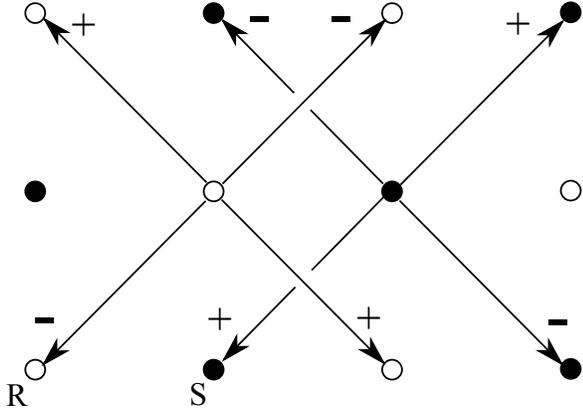}
\caption{Hopping terms that lead to a chiral mass $H_{C}$. The phase factors of ($\pm 1$) are labeled in the figure; the phase of a hopping term changes sign when rotated by $\pi/4$ and between sublattices $R$ and $S$ for the same direction. These phases correspond to a virtual magnetic flux of $\pi$ per plaquette (half a flux quantum). Since this virtual flux was already introduced to create the Dirac fermions, a chiral mass could be introduced simply through allowing next nearest neighbor hopping.
}\label{chiralhop}
\end{figure}

\begin{figure}
\includegraphics[width=3in]{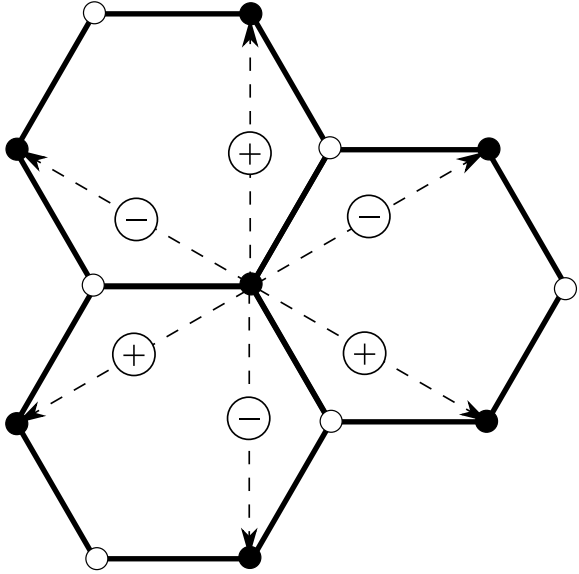}
\caption{The signs of the hopping amplitudes in Eq. (\ref{mcterm}) for a given sublattice in the hexagonal lattice. All signs are reversed for hopping within the other sublattice. Properly implemented, this term yeilds a chiral mass and will create gaps in the spectra of the 3 BEC fields.}\label{oddhop}
\end{figure}

\subsubsection{Mass terms}

In 2+1 dimensions, there are two linearly independent, Lorentz-invariant terms which lead to a massive fermion dispersion relation $\omega = \sqrt{v_{F}^{2} \bk^{2} + m_{0}^{2} + m_{c}^{2} }$,
\begin{eqnarray}\label{masses}
H_{m} = m \sum_\bk \psibar_\bk \psi_\bk, \\ H_{C} = i m_{C} \sum_\bk \psibar_{\bk} C_{1} C_{2} \psi_{\bk}.\nonumber
\end{eqnarray}
The first, $H_m$, breaks chiral symmetry, and can be experimentally engineered by introducing a superlattice potential.
 In terms of the original fermionic sublattices $R$ and $S$, we can write $H_m$ as
%
\begin{eqnarray}\label{basicmass}
H_{m} = m \sqof{ \sum_{\bx \in L_{S}} \psi_{S \bx}^{\dagger} \psi_{S \bx} - \sum_{\bx \in L_{R}} \psi_{R \bx}^{\dagger} \psi_{R \bx} }.
\end{eqnarray}
This term leaves the photon modes qualitatively unchanged, and as explained in Sec.~\ref{ssb} generates a density wave.

The chiral mass term $H_C$ is more interesting, as it has dramatic consequences for the photons.  As explained below, it can be engineered by introducing longer range fermion hopping terms with appropriately chosen phases.
%

%

In the presence of chiral-massive fermions, the resummation of 1-loop corrections to the photon propagator introduces a Chern-Simons term \cite{deserreview} in the action of the form
%
\begin{eqnarray}\label{mcs}
L_{CS} = \frac{\Delta}{4}  A_{\beta} F_{\alpha \gamma} \epsilon_{\alpha \beta \gamma},
\end{eqnarray}
with $\Delta\sim m_C$.
This term is not found in 3+1 dimensional theories.  Within the Feynman gauge it yields
a modified photon propagator:
\begin{eqnarray}\label{chiralprop}
D_{A \mu \nu} = \frac{-1}{p^{2} + \abs{\Delta}^{2}} \of{\delta_{\mu \nu} - \frac{p_{\mu} p_{\nu} - i \Delta \epsilon_{\mu \nu \alpha} p_{\alpha}}{p^{2}} } - \frac{p_{\mu} p_{\nu}}{p^{4}} \nonumber \\
 \,
\end{eqnarray}
with a photon mass gap
$\abs{\Delta}$.
The action $L_{CS}$ is only gauge invariant up to a boundary term, reducing the symmetry of the system to a subset of all possible gauge transformations.  The properties of QED3 with a chiral mass term are therefore sensitive to the topology of the 2+1 dimensional space.
 

In terms of the two sublattice fermion operators, the chiral mass term $H_{C}$ is
\begin{eqnarray}\label{chiralmassexp}
H_{C} &=& m_{C} \sum_{\bp} \left( \psi_{R \bK_{+} + \bp }^{\dagger} \psi_{R \bK_{+} + \bp} - \psi_{S \bK_{+} + \bp}^{\dagger} \psi_{S \bK_{+} + \bp}  \right.  \nonumber\\&&
  \left. -\psi_{R \bK_{-} + \bp}^{\dagger} \psi_{R \bK_{-} + \bp} + \psi_{S \bK_{-} + \bp}^{\dagger} \psi_{S \bK_{-} + \bp} \right). 
\end{eqnarray}
$H_{C}$ is invariant under a transformation which simultaneously switches the two sublattices and rotates between the two $\bK$ points, but is not invariant under either transformation individually. The simplest perturbation which obeys this symmetry is is a next-nearest-neighbor hopping term with direction-dependent phases, shown in figure \ref{chiralhop}. The Hamiltonian for this term is:
\begin{eqnarray}\label{CMashop}
H_{NNN} = \frac{m_{C}}{4} \sum_{\bx,j} \of{-1}^{x+y} \of{-1}^{j} \psi_{R/S \bx}^{\dagger} \psi_{R/S \bx + \mathbf{s}_{j} }
\end{eqnarray}
where the four $\mathbf{s}_{j}$ vectors join a site and its next nearest neighbors. The phases of these hopping terms change sign under a $\pi/4$ rotation and between the two sublattices $R$ and $S$. Such hopping phases are natural in our model, Eq.~(\ref{Htotal}), where the fermions experience an effective magnetic flux of $\pi$ per plaquette: the phases correspond to a flux of $\pi/2$ through each triangular half-plaquette. To lowest order in $\bp$, $H_{NNN}$ reduces to $H_{C}$. Tuning the lattice to allow nearest neighbor hopping would therefore introduce a chiral mass gap into the model. 

Engineering this chiral mass in our optical lattice would allow us to study Maxwell-Chern-Simons electrodynamics \cite{deserreview,appelquist1,appelquist2,coleman,hoshino,pimentel}. As seen in Eq.~(\ref{chiralprop}), there will be a gap in the photon spectrum.
Since the photon in our model can be identified with excitations about a BEC, this photon mass  implies the existence of a neutral Bose condensate with a gapped spectrum.  This unusual spectral feature
could be detected through RF spectroscopy \cite{campbell} or through its thermodynamic consequences.

Further, given that $p$ in Eq.~(\ref{chiralmassexp}) is small compared to ${\bf K_\pm}$, one can write, in terms of the $d$-density wave operator $\Theta$:
\begin{equation}
H_{c} \simeq 2 m_{c} \sum_{R,S} \psi_{R/S \bk + \cuof{\pi/l,\pi/l} }^{\dagger} \psi_{R/S \bk} \sin 2 \theta_\bk=2m_c \Theta
\end{equation}
where $\tan(\theta_\bk)=k_y/k_x$. Several authors, including Chakravarty \textit{et al} \cite{ddw},
have proposed that such \textit{d}-density wave order plays a role in
the  pseudogap phase of high temperature superconductors. \\[0.1in]

\centerline{\em Connection to the Haldane Model}
\vspace{0.1in}

On the hexagonal lattice the  hopping term, analogous to Eq.~(\ref{CMashop}),  leading to a chiral mass is
(Fig. \ref{oddhop}):
\begin{eqnarray}\label{mcterm}
 - i \frac{m_{C}}{4} \sum_{k , j=1}^{j=6} \of{-1}^{j} \of{ \psi_{R \mathbf{r}_{k}}^{\dagger} \psi_{R \mathbf{r}_{k}+\mathbf{r}_{j} } - \psi_{S \mathbf{r}_{k}}^{\dagger} \psi_{S \mathbf{r}_{k}+\mathbf{r}_{j} } }.
\end{eqnarray}
The sum on $k$ runs over all lattice sites in a given sublattice and $\mathbf{r}_{j}$ are the six vectors which join a site and its next nearest neighbors. This construction is analogous to the Haldane model \cite{haldane} for realizing conductance quantization in the absence of an external magnetic field in a hexagonal lattice. The contribution to the low energy theory from this mass is the same as in the square lattice.
In an experiment the appropriate phases could be engineered through the use of light assisted hopping, allowing the hopping elements to pick up local phases from an external laser beam \cite{jaksch}.

\subsubsection{Robustness}

An important issue  with our proposal
is that of stability: how robust is our construction against symmetry breaking perturbations? Inevitably, any realization of QED through this or a similar scheme must contend with anisotropies in the couplings, unequal or anisotropic Fermi and Bose velocities, and residual interactions which violate our emergent gauge invariance. Further, we have ignored the effect of the harmonic trapping potential required to confine the atoms; the long-ranged forces of QED could be disrupted if the trap were not shallow. 

These are difficult issues, and we are unable at this point to quantify all of them.  Each of these effects would have to be treated in detail, and their importance will inevitably hinge on technical details in the experiment.  Some particular perturbations have been studied.  For example, 
we have already seen that at low temperatures, QED3 is unstable towards spontaneous chiral symmetry breaking.  Near the transition temperature, small perturbations which break chiral symmetry can be quite important.  Other perturbations are less destructive: for example,  Vafek et al. \cite{vafek}, showed that anisotropy in the Fermi velocity is an RG-irrelevant perturbation at low energies.

%

We do have some additional knowledge of the robustness of this model agains Lorentz invariant perturbations.  
Through a simple power counting, we argue that any generic Lorentz invariant perturbation will have a vanishingly small contribution to the low energy physics of our system.  
For this argument, we work in units where the dimensionality [energy]=[momentum]=[length$]^{-1}$=[time$]^{-1}$ and label any quantity by the number of powers of energy that make it up.  For example, energy will have dimension +1, and length dimension -1. By noting that the action must be dimensionless,  the dimensionality of $\psi$ is 1, and $A$ is $1/2$.  According to the standard Wilsonian RG argument \cite{peskin}, only coupling constants with non-negative dimension will contribute to the low energy physics.  For example, mass terms and  the gauge coupling $e$ are relevant perturbations.
Along with the Chern-Simons term and the fermion masses, the terms in the QED3 action (\ref{LE}) are the only relevant and marginal operators consistent with Lorentz and gauge symmetries that can be constructed from $\psi$ and $A$ fields. 


\section{1+1 Dimensional Optical Lattice QED}\label{sec1d}
We now discuss the 1+1 dimensional analog to the model discussed in sections~\ref{sech2d} and \ref{secsli}.

\begin{figure}
\psfrag{u}{$\up$}
\psfrag{r}{$R \down$}
\psfrag{s}{$S \down$}
\psfrag{c1}{$\cuof{\psi,a_0}$}
\psfrag{c2}{$a_1$}
\includegraphics[width=3in]{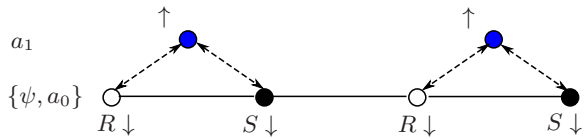}
\caption{Lattice for the 1d theory, with sublattices labeled by spin and by the particles that inhabit them. Bold lines represent ordinary hopping terms and dashed lines indicate hopping facilitated by optical transitions, as described in the text.}\label{1d}
\end{figure}

The  Feynman-gauge Hamiltonian for 1+1 dimensional QED (QED2) is
\begin{eqnarray}\label{H1dbare}
H =  \sum_k \psibar \gamma_1 \of{v_{F} k + e A_{1} } \psi + \frac{1}{2} \partial_\mu A_\nu \partial_\mu A_\nu 
\end{eqnarray}
where $\psi$ is a two component fermionic spinor, $A_\nu$ are bosonic fields, $\nu=0,1$, and $\gamma_1 = \sigma_y$. Following our previous arguments one would expect this to be emulated by the 1+1 dimensional generalizion of our planar Hamiltonian (\ref{Htotal}),
\begin{widetext}
\begin{eqnarray}\label{H1dstart}
& & \cuof{x \in L_{f}, y \in L_{b} } \nonumber \\
H &=& -t \sum_{x} \psi_{R/S\down x \pm \bl}^{\dagger} \psi_{S/R\down x}  + g_{0} \sum_x \psi_{R/S\down x}^{\dagger} \psi_{R/S\down x} n_{0 x} + g_1'' \sum_{x} \psi_{R/S\down x \pm \bl}^{\dagger} \psi_{S/R\down x} n_{1 \by = x \pm \bl/2}  \\ & &+ \sum_{b=0,1; k} \epsilon_{b k} n_{b k} + \sum_{b=0,1;x} \frac{V_b}{2} n_{b x/y} \of{n_{b x/y} - 1} + g_{1}' \sum_{<y x>} a_{1y + 2\bl}^{\dagger} a_{y} \of{\psi_{S x}^{\dagger} \psi_{S x} +\psi_{R x+\bl}^{\dagger} \psi_{R x+\bl} } + H.C. \nonumber \\
& & + g_{1}'' \sum_{<x y>} \psi_{R/S x}^{\dagger} \psi_{R/S x} n_{1 y = x \pm \bl/2}.
\end{eqnarray}
\end{widetext}
The lattice for this model is shown in figure \ref{1d}. Here, the $x$ represent fermion lattice sites ($L_f$), and the $y$ represent the boson lattice sites ($L_b$) associated with every other bond between fermion lattice sites. $\epsilon_{b k}$ is the boson kinetic term and chemical potential, and it has a minimum at $K = \pi/2l$. Bosons of species $a_0$ are confined to the fermion lattice sites and bosons of species $a_1$ occupy the $\by$ sites.

Unfortunately, the Hamiltonian in Eq. (\ref{H1dstart}) does not yield QED2. While this Hamiltonian trivially has fermions with a Dirac dispersion, 
the boson mediated fermion hopping term does not lead to a Bose-Fermi coupling of the form in Eq.~(\ref{H1dbare}).  Instead, the resulting coupling, $e \psibar A_1 \gamma_2 \psi$, has an erroneous extra phase factor of $e^{i \pi/2}$, and violates Lorentz Invarience. 
This same phase factor was also found in the 2D case, but there it was benign: it simply meant that $A_x$ is produced by bosons on the $y$-bonds, and $A_y$ is produced by bosons on the $x$-bonds. 
%
In the 1d case, however, we cannot make such a relabeling (since no $\gamma_2$ matrix appears in the fermion kinetic term). We can fix Eq. (\ref{H1dstart}) by implementing additional phase factors into the mediated hopping matrix elements. Conceptually, the best way to implement these phases is to work in a larger space, allowing the fermions to hop onto the $y$ sites, with a total energy cost of $U$ relative to the $x$ sites.  Using the techniques from \cite{spielman2,liusquare,zhu,jaksch}, one can engineer a spatially dependent phase on the hopping matrix elements, and produce a Hamiltonian,
\begin{widetext}
\begin{eqnarray}
& & \cuof{x \in L_{f}, y \in L_{b} } \nonumber \\\label{H1dbasic}
H &=& -t \sum_{x} \psi_{R/S\down x \pm \bl}^{\dagger} \psi_{S/R\down x} + U \sum_{x} \psi_{\up y}^{\dagger} \psi_{\up y} + g_{0} \sum_x \psi_{R/S\down x}^{\dagger} \psi_{R/S\down x} n_{0 x} + g_1 \sum_x \psi_{\up y}^{\dagger} \psi_{\up y} n_{1 y}  \\\nonumber & &+ \sum_{b=0,1; k} \epsilon_{b k} n_{b k} + \sum_{b=0,1;x} \frac{V_b}{2} n_{b x/y} \of{n_{b x/y} - 1} + g_{1}' \sum_{<y x>} a_{1y + 2\bl}^{\dagger} a_{y} \of{\psi_{S x}^{\dagger} \psi_{S x} +\psi_{R x+\bl}^{\dagger} \psi_{R x+\bl} } + H.C.+H_{\rm med}\\
H_{\rm med}&=& \omega_t  \sum_{<x y>} e^{i Q x} \psi_{\up y}^{\dagger} \psi_{R/S\down x}  + H.C.\label{hmed}
\end{eqnarray}
Here $\omega_t e^{i Q x}$ is the matrix element for a fermion to hop onto a boson site, with $Q=\pi/2l$,
and $U$ is the on-site energy.
Working to lowest nontrivial order in $\omega_t/U$, we integrate out the $y$-sites, obtaining 
\begin{eqnarray}\label{Ht}
H_{med} = \frac{\omega_{t}^{2}}{U} \sum_{x} \sqof{ \of{ \psi_{R x+\bl}^{\dagger} \psi_{S x} e^{i Q x} + \psi_{S x}^{\dagger} \psi_{R x+\bl} e^{-i Q x} } \of{1 + \frac{g_{1}}{U} n_{1 y}} +   \psi_{R/S x}^{\dagger} \psi_{R/S}  \of{1 + \frac{g_{1}}{U}  n_{1 y}} },
\end{eqnarray}
\end{widetext}
which gives the desired low energy Hamiltonian.

\section{3+1 Dimensions}

Full 3+1 dimensional QED can be simulated by combining the schemes which we have presented for one and two dimensions. We will work with a cubic lattice formed by layering the 2D square lattice of section III, with lattice spacing $l_z$ between each layer. Each 2D layer is the same as the lattice of section III, containing three bosons and one fermion distributed between the various sublattices. Between the layers we add a fourth boson species $a_3$, which occupies the space between every other layer and can only interact with the fermions through a light-assisted transition as in the 1d case. We shall then implement additional phases to change the sign of the hopping amplitudes $t_z$ between planes; we choose the phases such that vertical hopping between sites on the $R$ sublattice gets no additional phase but $z$-directional hopping matrix elements between sites on the $S$ sublattices are opposite in sign. We ignore any vertical hopping beyond nearest neighbors. Under those conditions, our free particle Hamiltonian is
\begin{eqnarray}
H &=& -2 \sum_\bk \psi_{\bk}^{\dagger} M \psi_\bk, \\
M &=& \of{ \begin{array}{cc} t_z \cos k_z l_z & t \cos k_x l + i t \cos k_y l \\ t \cos k_x l - i t \cos k_y l & -t_z \cos k_z l_z \end{array}}. \nonumber
\end{eqnarray}
This Hamiltonian exhibits eight Dirac points, located at $\bk = \pi \cuof{\pm 1/2l, \pm 1/2l, \pm 1/2 l_z}$, though many of these are degenerate. We can obtain the necessary bispinor representation by simply combining pairs of them; for example, the combination of $\bK = \pi \cuof{ 1/2l, 1/2l, 1/2 l_z }$ and $-\bK$ yeilds the following representation for the Lorentz group
\begin{eqnarray}
\Gamma_0 = \of{ \begin{array}{cc} 0 & 1_{2\times2} \\ 1_{2\times2} & 0 \end{array}}, \Gamma_j = i \of{ \begin{array}{cc} 0 & \sigma_j \\ - \sigma_j & 0 \end{array}}.
\end{eqnarray}
The $A_z$ component of the gauge field is derived from the $a_3$ bosons in the same manner as in section~\ref{sec1d}. Other than ensuring that the sign flips between sublattices are preserved when adding the gauge interaction, the derivation is essentially unchanged.

\section{Summary and Outlook}\label{seccon}


We have proposed a number of experiments involving neutral atoms in optical lattices which will emulate the properties of relativistic quantum electrodynamics. These models are technically very difficult to realize, requiring a large number of lasers and atomic species that all must be fine-tuned to produce the desired behavior. However, the necessary technologies (optically induced phases, interpenetrating spin dependent lattices, and tunable Bose-Fermi mixtures) for our model have all been separately realized in recent experiments. Further, as a weakly coupled route from a lattice model to a Lorentz-invariant dynamical gauge theory, the scheme we have proposed here is also of significant theoretical interest.

Finally, it would be interesting to establish whether or not this scheme could be extended to simulate more complicated gauge theories, such as non-abelian theories or perturbative gravity. Given the importance of these theories in high energy physics, a method for simulating them in cold atoms would be a very significant development. The difficulty of this process should not be underestimated. These theories intrinsically have a very large number of degrees of freedom; in 2+1 dimensions, an SU(2) theory has nine gauge boson modes (three components of spacetime with three SU(2) generators associated with each), and an SU(3) theory has twenty-four! Further, non-abelian gauge theories have boson self-interactions that must be finely tuned to avoid breaking gauge invariance.  Worse yet, the gauge fields exhibit ghostlike degrees of freedom with negative norm. Accounting for all of these effects in any scheme would be a daunting. Nonetheless, quantum simulation of non-abelian dynamical gauge theories is a fascinating prospect, and the theory we have described here could be an important first step in that direction.

\section{Acknowledgments}

Eliot Kapit would like to thank Andre LeClair, Liam Mcallister and David Marsh for illuminating discussions related to this work. This work was supported by NSF grant PHY-0758104, a grant from the Army Research Office with funding from the DARPA OLE program, and by the Department of Defense (DoD) through the National Defense Science \& Engineering Graduate (NDSEG) Program.

\end{document}